\documentclass[aps,twocolumn,showpacs, superscriptaddress,amsmath,amssymb]{revtex4-1}
\usepackage{graphicx}
\usepackage{epsfig}
\usepackage{epstopdf}
\usepackage{bm}
\usepackage[usenames,dvipsnames]{color}
\usepackage{float}

\begin{document}
\title{Origin of Suppression of Proximity Induced Superconductivity in Bi/Bi$_2$Sr$_2$CaCu$_2$O$_{8+\delta}$ Heterostructure}
\author{Asish K. Kundu}
\email{asishkumar2008@gmail.com}
\author{Ze-Bin Wu}
\author{I. K. Drozdov}
\author{G. D. Gu}
\author{T. Valla}
\email{valla@bnl.gov}
\affiliation{Condensed Matter Physics and Materials Science Department, Brookhaven National Lab, Upton, New York 11973, USA}

\begin{abstract}
Mixing of topological states with superconductivity could result in topological superconductivity with the elusive Majorana fermions potentially applicable in fault-tolerant quantum computing. One possible candidate considered for realization of topological superconductivity  is thin bismuth films on Bi$_2$Sr$_2$CaCu$_2$O$_{8+\delta}$ (Bi2212). Here, we present angle-resolved and core-level photoemission spectroscopy studies of thin Bi films grown {\it in-situ} on as-grown Bi2212 that show the absence of proximity effect. We find that the electron transfer from the film to the substrate and the resulting severe underdoping of Bi2212 at the interface is a likely origin for the absence of proximity effect. We also propose a possible way of preventing a total loss of proximity effect in this system. Our results offer a better and more universal understanding of the film/cuprate interface and resolve many issues related to the proximity effect.

\end{abstract}


\maketitle

\section{Introduction}
Topological superconductors (TSCs) are the class of materials where the  Majorana fermions are expected to exist, potentially playing an important role in the future developments of quantum computing \cite{hasan2010colloquium,ando2015topological,qi2011topological,alicea2012new}. The research in this area has been expanding quickly after the theoretical predictions of topological nature in 2D {\it p+ip}-wave \cite{read2000paired} and 1D \textit{p}-wave \cite{kitaev2001unpaired} superconductors, where the Majorana bound states are predicted to emerge at the vortices and ends of wire, respectively. However, due to the lack of intrinsic \textit{p}-wave superconductors and their expected very low transition temperature and gap-value, the experimental detection of Majorana modes is lacking. Apart from the intrinsic TSCs, another option, where the superconductivity would be induced at the interface of a superconductor and topological material by tunneling of Cooper-pairs across the interface has been also explored. Known as the superconducting proximity effect, pioneered by de Gennes {\it et al.} \cite{de1964boundary} in the 1960s, the phenomenon   has been proven very promising in achieving topological superconductivity in topological insulator (TI) films grown on superconducting (SC) substrates, such as NbSe$_2$ \cite{wang2012coexistence,xu2014momentum,xu2014artificial,xu2015experimental,sun2016majorana}, FeTe$_{0.55}$Se$_{0.45}$ \cite{chen2018superconductivity} and elemental SCs, such as Pb and Sn \cite{qu2012strong,yang2012proximity}. Again, all these systems involve low-$T_c$ superconductors with small gaps, making the detection of Majorana modes extremely difficult.

Cuprate superconductors might seem as ideal candidates to realize TSC, as they have an order of magnitude larger gaps and higher transition temperatures. However, it is not clear if the proximity effect would be efficient at the interfaces involving materials with incompatible crystal and pairing symmetries \cite{qi2011topological}, although some theoretical studies suggest that the proximity effect may even be enhanced by the mismatch of the TI films and the cuprate substrates \cite{li2015realizing}. Several experimental attempts have been made on TI/Bi2212 heterostructures with the conflicting results \cite{zareapour2012proximity,wang2013fully,yilmaz2014absence,xu2014fermi}. A gap-like features at the Fermi level were reported for Bi$_2$Se$_3$/Bi2212 system in scanning tunneling microscopy (STM) \cite{zareapour2012proximity,wan2019twofold} and angle-resolved photoemission spectroscopy (ARPES) \cite{wang2013fully} studies, interpreted as a strong evidence of proximity-induced superconductivity in the topological material. These results were later re-examined by similar studies showing the absence of the superconducting gaps \cite{yilmaz2014absence,xu2014fermi}. In a very recent STM study by Rachmilowitz {\it et al.} \cite{rachmilowitz2020coulomb}, authors claim that the observed gap in the tunneling spectra (dI/dV) is not related to superconductivity, but is induced by a Coulomb blockade.  Theoretical calculations also suggest that the $s$-wave gap claimed on Bi$_2$Se$_3$/Bi2212 system by Wang {\it et al.} \cite{wang2013fully} may not be purely superconducting in origin \cite{li2016theoretical}.
In a recent experimental study, Shimamura {\it et al.} \cite{shimamura2018ultrathin} suggest that the TI is not required for the interface TSC and that the later may be realized even in Bi/Bi22212 heterostructures by inducing the pairing in the Rashba-spin-orbit-coupled states of Bi. By combining  ARPES and STM study, they concluded that the observed gap in the Bi-states is due to the proximity induced superconductivity. Obviously, all these contradictory results make the situation very complicated, with the fundamental question whether the cuprate SCs  (in particular Bi2212) can induce superconductivity in the films still remaining open.
\begin{figure*}[ht!]
\centering
\includegraphics[width=12cm]{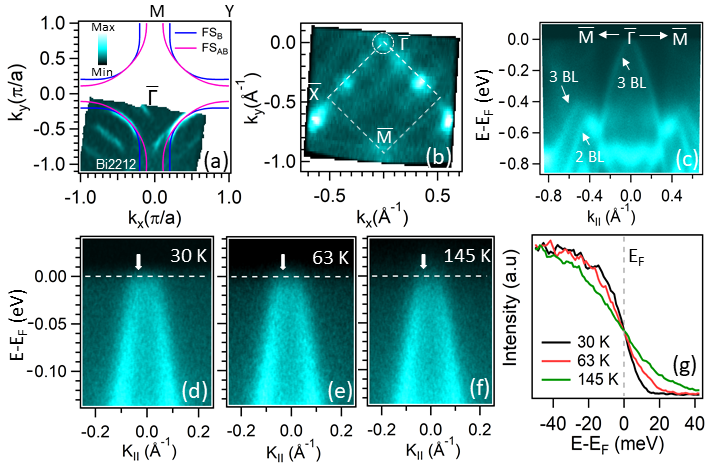}
\caption {Electronic structure of as-grown Bi2212 (OD91, $T_c$=91 K) and Bi films grown on it. (a) Fermi surface (FS) map of Bi2212 at 30~K with $h\nu$=21.2 eV, where blue (magenta) lines represent the bonding (antibonding) states. (b) FS of $\sim$ 9{\AA}~Bi/Bi2212. The dotted square represents one quarter of Bi(110) Brillouin zone. (c) Band dispersions along the $\bar{\Gamma}$-$\bar{M}$ of Bi(110) at 30~K. (d)-(f) Zoomed in views near the Fermi energy region of (c) at various temperatures, as indicated. (g) Energy distribution curves (EDCs) at the momenta indicated by arrows in figures (d)-(f).}\label{Fig1}
\end{figure*}

Here, we have studied the valence band and core-level electronic structure of {\it in-situ} grown Bi films on as-grown and overdoped Bi2212. We find no evidence of proximity induced gap at the FS formed by Bi-derived states. Our core-level measurements show that the Bi{\it5d} states of Bi2212 shift to higher binding energies upon Bi deposition, indicating  a severe underdoping at the interface. In the case of as-grown Bi2212, the electron transfer from the film to the substrate results in a complete loss of superconductivity at the interface, representing the fundamental obstacle for the proximity effect. For the sub-monolayer coverages, the loss of hole doping is also evident from the Luttinger count (Fermi surface volume). Our results also indicate that the complete loss of superconductivity can be avoided if the films are grown on heavily overdoped Bi2212.

\section{Experimental results and discussions}
The bulk bismuth crystallizes in the A7 rhombohedral structure. In case of thin films, it is know that the hexagonal phase of bismuth Bi(111) is preferred over the Bi(110) phase for the thicker films {\cite{shimamura2018ultrathin}}. The Bi(110) structure is stabilized in the thinner films, while the transition to Bi(111) occurs at a critical thickness that is dictated by the substrate \cite{bian2014first}. In the present study  the film thickness was $\le$~9~{\AA}, with only the Bi(110) structure observed. The thickness was calibrated by measuring the core-level attenuation of Bi 5$d$$_{5/2}$ peak of Bi2212 and sometimes expressed in units of bilayer (BL) corresponding roughly to $\sim$3.3~{\AA} of Bi.

\begin{figure*}[ht!]
\centering
\includegraphics[width=14cm]{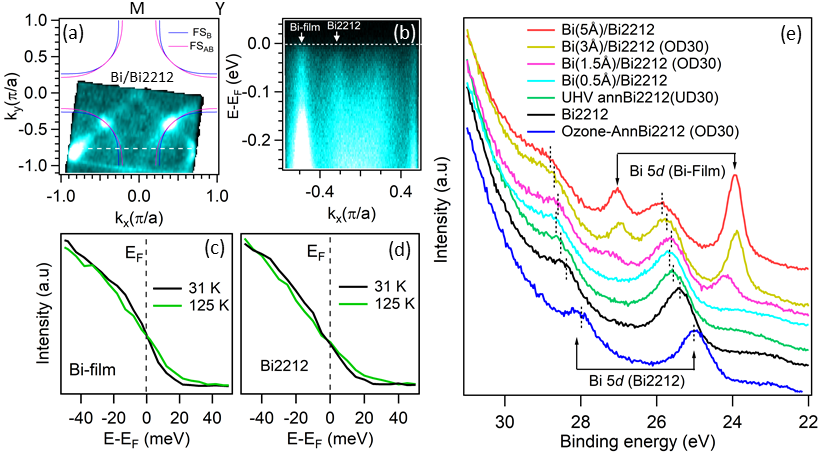}
\caption { Electronic structure of $\sim$ 5{\AA}~Bi/Bi2212. (a) FS. Contribution of states from the Bi2212 and Bi(110) are clearly visible. Blue (magenta) lines represent the bonding (antibonding) states. (b) Band dispersions along the cut near the antinode as indicated by dotted line in (a). (c), (d) EDCs at different temperatures for the Bi and Bi2212 states, respectively at the momenta as indicated by arrows in (b). (e) Core-levels of Bi 5{\it d} for different coverages of Bi on Bi2212 plotted together with the as-grown Bi2212 (OD91), UHV annealed Bi2212 (UD30) and the ozone annealed Bi2212 (OD30). Films grown on as-grown and ozone annealed Bi2212 substrates are abbreviated as `Bi/Bi2212' and `Bi/Bi2212(OD30)', respectively. Changes of Bi 5{\it d} peak positions related to Bi2212 are marked by vertical bars.}\label{Fig2}
\end{figure*}
Figure \ref{Fig1} shows the electronic structure of an as-grown Bi2212 substrate (OD91, $T_c$ = 91 K) and of a Bi film ($\sim$ 9~{\AA}) grown on top of it. The FS of the bare Bi2212 substrate is shown  in figure \ref{Fig1}(a). Due to the presence of the $d$-wave gap, only the near-nodal segments are visible, while the intensity is strongly reduced in most of the Brillouin zone (BZ).  The blue (magenta) lines represent the bonding (antibonding) states obtained from the fitting of the FS using the tight-binding formula that best fits the experimental data \cite{drozdov2018phase,Valla2020}. Figure \ref{Fig1}(b) shows the intensity at the Fermi level after depositing a 9~{\AA} thick Bi film on Bi2212. It can be seen that the Fermi surface has  4-fold symmetric features, with a hole pocket at the zone center ($\bar{\Gamma}$ point) developed upon Bi deposition. The similar hole pocket is also observed on the Bi(110) surface of a bulk single crystal due to the formation of surface states in the projected bulk band-gap \cite{hofmann2006surfaces}. We also observe several new features that were not observed on the Bi(110) surface of a bulk crystal \cite{hofmann2006surfaces}. These extra features form nearly one-dimensional lines at the FS, running perpendicular to each other and likely correspond to the surface/interface related states. Another feature with strong intensity is observed along $\bar{\Gamma}$-$\bar{X}$, close to the nodal position of the Bi2212 substrate. A very complex electronic structure of the Bi layers on Bi2212 is not  surprising and is in line with previous studies  \cite{koroteev2008first}. The dispersion of states along the $\bar{\Gamma}$-$\bar{M}$ for the present film is shown in figure \ref{Fig1}(c). A comparison with the reported theoretical calculations \cite{shimamura2018ultrathin,bian2014first,koroteev2008first} suggests that the observed states correspond to the 2 and 3 BL Bi(110) films.

We now turn to the Bi-derived states near the Fermi level to determine if there is any proximity-induced gap opening. Figures \ref{Fig1}(d)-(f) are the zoomed in views of the figure~\ref{Fig1}(c) near the Fermi level recorded at various temperatures, 30~K, 63~K, and 145~K, respectively. The energy distribution curves (EDCs) at the momenta indicated by arrows in figures~\ref{Fig1}(d)-(f) are plotted together in figure~\ref{Fig1}(g). We do not observe any leading edge shift of EDCs with temperature, indicating an absence of proximity effect in the present film. This is in sharp contrast to the recent study by Shimamura {\it et al.} \cite{shimamura2018ultrathin} who have reported a gap opening in the Bi films of similar thickness. We might expect that the thicker the film, the less chance for inducing the pairing at the surface of the film, as the proximity effect is generally limited by the $c$-axis coherence length. The $c$-axis coherence length of as-grown Bi2212 is very short ($\sim$ 1~{\AA}) \cite{Kang1988,Kim1999} in comparison to the Bi film thickness $\sim$ 9~{\AA}, and could by itself be the reason for the absence of  proximity induced superconductivity at the surface of the film \cite{yilmaz2014absence}. However, if this is the fundamental limitation, with the reduced film thickness, and in particular, for the sub-monolayer coverage, one would expect the superconductivity to survive in the surface region probed by ARPES. Therefore, we have performed a very careful analysis of thinner films where simultaneous monitoring of the states arising both from the substrate and a Bi-film could be performed to better understand what is happening at the Bi-Bi2212 interface.

Figure~\ref{Fig2}(a) displays the FS of a somewhat thinner $\sim$ 5~{\AA} Bi film on as-grown Bi2212 where the photoemission signal originating from both the substrate and the film can be identified. Figure~\ref{Fig2}(b) shows the dispersion of the states near the antinodal region of Bi2212 along the dotted-line indicated in figure~\ref{Fig2}(a). The substrate and the Bi-film related features are marked. Temperature dependent EDCs at the momenta indicated in figure~\ref{Fig2}(b) for the Bi-films and substrate are presented in figure~\ref{Fig2}(c) and (d), respectively. Remarkably, with temperature variation, we do not observe any gaps (or leading-edge shifts) in EDCs not only in Bi-derived states but also in underlying Bi2212 substrate related states. Furthermore, the quasiparticle peak is also totally absent in the Bi2212 state. This clearly indicates that the superconductivity is strongly suppressed or totally absent at the interface, even in the Bi2212 states.
\begin{figure*}[ht!]
\centering
\includegraphics[width=12cm]{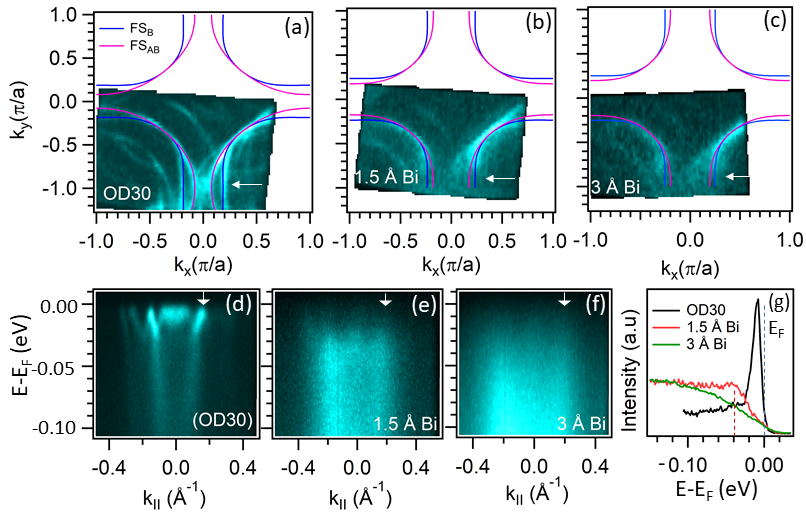}
\caption { Electronic structure evolution showing transition from overdoped to underdoped Bi2212 upon Bi deposition. (a)-(c) FS of a bare overdoped Bi2212(OD30), as covered by 1.5{\AA}~Bi and 3{\AA}~Bi, respectively. (d)-(f) Band dispersions along the cut (indicated by arrows in (a)-(c)) near antinode. (g) EDCs along the momentum cut as indicated by arrows in (d)-(f).}\label{Fig3}
\end{figure*}
To clarify the origin of this loss, we look at the spectra of the Bi 5{\it d} core-levels of Bi/Bi2212 for several film thicknesses and compare them to the Bi 5{\it d} spectra corresponding to the bare Bi2212 at different doping levels. The binding energy of the Bi 5{\it d} levels correlates with, and can serve as a precise measure of the hole doping of a Bi2212 surface. Figure~\ref{Fig2}(e) shows the Bi 5{\it d} levels for several characteristic doping levels of bare Bi2212 and several Bi films grown on as-grown and overdoped Bi2212. It can be seen that the Bi 5{\it d} core-levels of Bi2212 shift towards higher binding energy upon deposition of Bi. We note that this shift saturates beyond the Bi thickness of $\sim$ 1-2 {\AA} and  $\sim$ 4 {\AA} on as-grown and overdoped Bi2212, respectively. Interestingly, for all film thicknesses studied here, this energy shift is larger than the one corresponding to the underdoping of an as-grown Bi2212 ($T_c=91$ K) to UD30 ($T_c$=30 K) by vacuum annealing \cite{drozdov2018phase}. Thus, our result indicate that there is a severe underdoping at the Bi2212 interface upon Bi deposition due to the electron transfer from the film to the substrate. This results in a loss of hole doping (underdoping) at the interface, going beyond the superconducting phase. Indeed our tight-binding fitting of the FS reveals a doping level of probed Bi2212  of around 0.01$\pm$0.015, indicating a \lq\lq{}dead\rq\rq{} substrate, consistent with the the core-level data. This explains the absence of superconducitvity  (a gap and a quasiparticle peak) in both the Bi2212 and film-derived states. Therefore, we have identified the charge transfer from the film to the substrate as the main reason for the observed loss of proximity in this system that is detrimental even at the sub-monolayer film thickness, where the c-axis coherence length would not be a limiting factor. In fact, that same effect has been already used by Zhang \textit{et al.} \cite{Zhang2016} to reduce the doping level of Bi2212 by depositing small amounts of potassium on the cleaved surface. We have also tested if deposition of Pd causes similar effects. Indeed, the situation seems universal: deposition of a metal on the surface of Bi2212 induces the charge transfer and resulting underdoping of the Bi2212 in the interface (surface) region that effectively destroys superconducitvity.

Then the question may arises whether the cuprates are totally useless in inducing superconductivity. A possible way of preventing the total loss of superconductivity could be to start with the heavily overdoped Bi2212 sample as a substrate. The hope is that after Bi deposition the overdoped Bi2212 would lose some hole doping, but would still remain superconducting. To check this, we used the heavily overdoped Bi2212 (OD30, $T_c$=30 K), obtained by ozone annealing as a substrate for Bi films.

The electronic structure of a heavily overdoped Bi2212 (OD30) and its evolution with the Bi deposition are presented in figure~\ref{Fig3}. The top and the bottom rows show evolution of the FS and corresponding band dispersions near the antinodal cut, respectively. The doping level of the Bi2212 substrate before and after deposition of Bi is calculated from the FS area. The obtained doping levels are 0.27, 0.09 and 0.04 for the bare overdoped Bi2212 and for 1.5  and 3 {\AA} Bi films, respectively. Up to a nominal Bi coverage of $\sim$ 3 {\AA} we do not see any additional states at the FS other than those corresponding to  Bi2212 substrate. It could be that the Bi islands are still too small and unconnected to form Bi-related states. However, Bi-deposition has a significant effect on the photoemission intensities near the antinodal region of the FS of Bi2212. It can be seen that after Bi deposition, the intensities at the antinode are strongly suppressed (figure~\ref{Fig3}(b) and (c)). These effects can be better visualized from the near antinodal cuts (figure~\ref{Fig3}(d)-(f)) as well as from the corresponding EDC plots (figure~\ref{Fig3}(g)). By comparing figure~\ref{Fig3}(d) and (e) it is clear that the Bi deposition shifts the Bi2212 states far below the Fermi level resulting in suppression of density of states near the E$_F$. The EDCs in figure~\ref{Fig3}(g) shows that upon deposition of 1.5 {\AA} Bi on Bi2212, the quasiparticle peak is shifted from 7 to 40 meV. Further deposition of Bi (3 {\AA}) completely destroys the superconductivity as evident from the absence of quasiparticle peak, consitent with the doping level of p $\sim$0.04 obtained from the FS.

As already noted, up to the film thickness of 3 {\AA}, no Bi-derived states are observed at the FS.  However, the presence of Bi overlayer and its evolution can be monitored by the Bi 5{\it d} core-levels, as shown in figure~\ref{Fig2}(e). With increasing film coverage, Bi 5{\it d} core-levels corresponding to the substrate shift towards the higher binding energy. This indicates a transformation from the overdoped to underdoped Bi2212 at the film/substrate interface. This result is fully consistent with the the FS and valence band evolution discussed above. It is also evident that the Bi/Bi2212 interface loses superconductivity faster in the as-grown samples than in  the overdoped ones.
\begin{figure}[ht!]
\centering
\includegraphics[width=8cm]{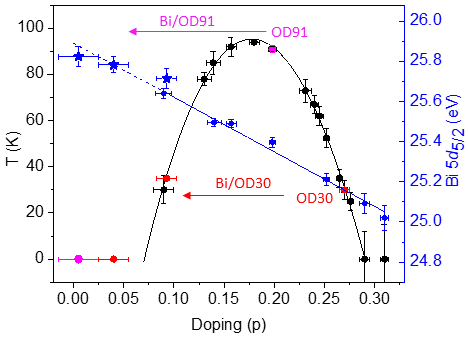}
\caption { Phase diagram of Bi2212 and its evolution with Bi deposition. The phase diagram of bare Bi2212 was obtained by annealing the film in ozone (overdoping), or UHV  (underdoping) and the doping levels (p) were determined from the fitting of the FS. The doping levels versus the Bi 5$d$$_{5/2}$ peak position and $T_c$ are plotted with blue and black filled-circles, respectively. Most of these data points are from Drozdov {\it et al.}, \cite{drozdov2018phase}. The solid-line (blue) is the linear fit of the Bi 5$d$$_{5/2}$ peak position with doping. The dotted blue-line is the extrapolation of this fitting. Position of the substrate Bi 5$d$$_{5/2}$ peak after deposition of various thickness of Bi films are shown by the asterisk symbols (blue). The change of doping levels and $T_c$ for the as-grown and overdoped substrates upon deposition of Bi is indicated in magenta and red circles with arrows, respectively.} \label{Fig4}
\end{figure}

To summarize our results, we have drawn the Bi2212 phase diagram showing the substrate Bi 5$d$$_{5/2}$ peak position and superconducting temperature ($T_c$) versus doping in figure~\ref{Fig4}. The position of the Bi 5$d$$_{5/2}$ peak is indicated for various doping levels of the bare Bi2212 samples, obtained by ozone (overdoping), or UHV annealing (underdoping). In each case, the doping level was obtained from the fitting of the FS \cite{drozdov2018phase} and the corresponding Bi 5$d$$_{5/2}$ peak position and $T_c$ are plotted with blue and black filled-circles, respectively. Most of these data points are from the same samples studied by Drozdov {\it et al.,} \cite{drozdov2018phase}.  The Bi 5$d$$_{5/2}$ peak positions were obtained by fitting the spectra using Lorentzian peak shape with a cubic background.
The Bi 5$d$$_{5/2}$ peak position varies nearly linearly with doping and can be fitted with a straight line (solid blue).  Similarly, after Bi deposition, the Bi 5$d$$_{5/2}$ peak position associated with the substrate is shown by the asterisk symbol (blue), at the corresponding reduced doping. Surprisingly, these data points also fall on the same straight line, within the error bar. After Bi-deposition, variation of $T_c$ of the substrate (at the interface region) can be obtained by projecting the calculated doping levels of Bi2212 to the superconducting dome. The change in $T_c$ and doping levels for the as-grown and overdoped substrates upon deposition of Bi is indicated in magenta and red circles and arrows, respectively. Projection of doping to the superconducting dome shows that the as-grown Bi2212 (OD91, $T_c$=91~K) becomes non-superconducting (UD0, $T_c$=0~K) upon deposition of 5 {\AA} of Bi. Similarly, strongly overdoped Bi2212 (OD30) substrate with $T_c$=30~K is transformed into an underdoped one (UD35) with $T_c$=35~K upon deposition of 1.5 {\AA} Bi and outside of the superconducting dome for 3 {\AA} Bi (UD0, $T_c$=0~K).

\section{Conclusion}
In summary, we have studied the electronic structure of Bi films of various thicknesses on Bi2212. Our results show no evidence of a proximity-induced gap opening in the film-derived states, in contrast to the recent study by Shimamuraet {\it et al.} \cite{shimamura2018ultrathin}. By depositing thinner films and accessing both the substrate and overlayer Bi-states, we discover the fundamental reason for the suppression of proximity effect. Both ARPES and core-level results demonstrate that the Bi deposition modifies the doping level, severely reducing the hole doping and causing the loss of superconductivity at the film-substrate interface. Our results suggest that this phenomenon is universal and that proximity effect will be suppressed for any metal deposited on as-grown Bi2212. Despite the negative results, this study offers a better understanding of the film - cuprate superconductor interface, helps in understanding contradictory results and offers the strategy for avoiding the total loss of superconductivity at the interface.

\section*{Experimental Section}
The experiments within this study were performed in an experimental facility that integrates oxide-molecular beam epitaxy (OMBE), ARPES, and STM in a common ultra high vacuum (UHV) system \cite{Kim2018a}. Bi films were grown {\it in-situ} on as-grown and overdoped Bi2212 substrates. The starting, as-grown Bi2212 crystals were synthesized by the traveling floating zone method, clamped to the sample holder and cleaved with Kapton tape in the ARPES preparation chamber. The overdoping is achieved by annealing the as-grown samples in ozone  \cite{drozdov2018phase,Valla2020}. High purity bismuth (99.99\%) was evaporated from a resistive evaporator with Al$_2$O$_3$ crucible. During the film growth, substrate was kept at room temperature (RT). The photoemission experiments were carried out on a Scienta SES-R4000 electron spectrometer with the monochromatized He I$_\alpha$(21.2 eV) and He II$_\alpha$(40.8 eV) radiation (VUV-5k). The total instrumental energy resolution was 8 and 20 meV for the ARPES and core-level measurements, respectively. Angular resolution was better than 0.15$^{\circ}$ and 0.4$^{\circ}$ along and perpendicular to the slit of the analyzer, respectively. Most of the data were taken at 30 K, except for the temperature dependence measurements.

\section*{Acknowledgements}
This work was supported by the US Department of Energy, office of Basic Energy Sciences, contract no. DE-SC0012704.

\section*{Conflict of Interest}
The authors declare no conﬂict of interest.

\bibliography{ref_Bi2212}

\end{document}